\documentclass[runningheads]{llncs}

\usepackage[T1]{fontenc}
\usepackage{amsmath,amssymb}
\usepackage{graphicx}
\usepackage{hyperref}
\begin{document}

\title{Localization Gains from Explanation Regularization Should Not Be Interpreted as Faithfulness Gains}
\titlerunning{Localization Is Not Faithfulness}

\author{Amrin Akter\inst{1} \and
Mauro Dalla Mura\inst{1} \and
Diana Mateus\inst{2} \and
Dawood Al Chanti\inst{1}}
\authorrunning{A. Akter et al.}
\institute{
Univ. Grenoble Alpes, CNRS, Grenoble INP, GIPSA-lab, Grenoble, France \\
\email{\{amrin.akter,mauro.dalla-mura,dawood.al-chanti\}@grenoble-inp.fr}
\and
Nantes Université, École Centrale Nantes, LS2N, Nantes, France \\
\email{diana.mateus@ls2n.fr}
}
\maketitle

\begin{abstract}
Training-time explanation regularization is often evaluated by whether saliency maps become more concentrated within annotated regions. We argue that this criterion alone is insufficient: transformation consistency, lesion localization, and faithfulness to the classifier are distinct properties and should be evaluated separately, particularly in medical imaging. We examine this distinction through a case study of Contrastive Grad-CAM Consistency (CGC) in three-class brain tumor MRI classification. Using a single-channel ResNet-50 with five-fold patient-grouped cross-validation, we compare a cross-entropy baseline with CGC. CGC changes overall Pointing Game (PG) from $27.24\%$ to $36.02\%$ and Relevance Mass Accuracy (RMA) from $11.17\%$ to $18.01\%$, while classification accuracy changes from $93.63\%$ to $94.87\%$. Perturbation-based faithfulness metrics show inconsistent changes: Insertion Area Under the Curve (IAUC) changes from $0.73$ to $0.78$, whereas Deletion Area Under the Curve (DAUC) favors the baseline ($0.36$ versus $0.41$; lower is better), and overall Iterative Removal Of Features (IROF) is $0.57$ for both models. These results support our position that improved spatial agreement with pathology should not be interpreted as a general improvement in explanation faithfulness. We therefore recommend property-specific evaluation in which the property targeted by the explanation objective, spatial agreement with reference regions, and model reliance are measured and reported separately.

\end{abstract}

\section{Introduction}
\label{sec:intro}

High predictive accuracy in medical imaging does not ensure that a classifier relies on clinically relevant image evidence. Saliency maps are commonly used to identify image regions associated with the model prediction. These explanations, however, require independent evaluation. A visually plausible heatmap does not establish that the highlighted regions contribute substantially to the model prediction~\cite{adebayo2018sanity,arun2021assessing,saporta2022benchmarking}.

This distinction becomes particularly important when explanations are explicitly regularized during training. Greater spatial concentration of Grad-CAM saliency within a lesion does not, by itself, establish greater interpretability or faithfulness. We therefore distinguish three properties:

\begin{enumerate}
    \item \textbf{Transformation consistency}: whether an explanation responds consistently to a label-preserving input transformation.
    \item \textbf{Localization}: whether saliency aligns spatially with a clinically or semantically meaningful reference region.
    \item \textbf{Faithfulness}: whether regions ranked as important affect the model output under a specified intervention.
\end{enumerate}

These properties are not interchangeable. A transformation-consistent heatmap may still consistently highlight shortcut features. A heatmap may overlap a tumor without capturing all image evidence used by the classifier. Conversely, a perturbation-based metric may assign a more favorable score to an explanation whose influential regions extend beyond the annotated lesion. Using one property as a proxy for another can therefore support claims that exceed the corresponding evidence.

We examine this position through a case study of Contrastive Grad-CAM Consistency (CGC)~\cite{pillai2022consistent} on the Brain Tumor MRI dataset~\cite{Cheng2017brain}. CGC explicitly encourages consistency between Grad-CAM maps under geometric transformations without requiring pixel-level explanation annotations. Using patient-grouped cross-validation, we compare CGC with a cross-entropy baseline in terms of lesion localization and perturbation-based faithfulness.

The case study provides empirical evidence for this distinction; it does not introduce a new XAI method or constitute clinical validation. CGC yields higher mean lesion-localization scores, whereas the three perturbation-based faithfulness metrics do not indicate a consistent faithfulness gain. This empirical separation motivates the following position:

\begin{quote}
\emph{Explanation-aware training should be evaluated using separate, explicitly named criteria for the property targeted by the objective, spatial agreement with reference regions, and model reliance. Improvement on one criterion should not be interpreted as improvement on the others.}
\end{quote}

The remainder of the paper is organized as follows. Section~\ref{sec:related} reviews explanation regularization and its evaluation, Section~\ref{sec:case} presents the CGC case study, and Section~\ref{sec:evidence} reports the empirical evidence. Section~\ref{sec:implications} discusses the implications for XAI evaluation, followed by the limitations in Section~\ref{sec:limitations} and the conclusion in Section~\ref{sec:conclusion}.

\section{Explanation Regularization and Its Evaluation}
\label{sec:related}

Post-hoc visual explanation methods highlight input regions associated with model predictions without modifying the trained model. Gradient-based methods derive feature importance from class-score gradients~\cite{simonyan2013deep} or path-integrated gradients~\cite{sundararajan2017axiomatic}. Class-activation methods construct spatial explanations from intermediate convolutional feature maps. Class Activation Mapping (CAM) combines these maps using classifier weights~\cite{zhou2016CAM}, whereas Grad-CAM uses class-specific gradients to determine their weights and applies to a broader range of convolutional neural network (CNN) architectures~\cite{selvaraju2017grad}. Subsequent variants modify the weighting or aggregation of the feature maps, including Grad-CAM++~\cite{chattopadhay2018gradcamp}, Score-CAM~\cite{wang2020score}, LayerCAM~\cite{jiang2021layercam}, and Eigen-CAM~\cite{banymuhammad2021eigencam}. Perturbation-based approaches estimate importance from changes in model predictions under input perturbations~\cite{fong2017meaningful,petsiuk2018rise}. Visually plausible saliency maps, however, do not necessarily provide reliable accounts of model behavior. Randomization tests show that some saliency methods are insensitive to model-parameter or label randomization~\cite{adebayo2018sanity}, while medical imaging studies report substantial variation among explanation methods and limited agreement with annotated pathology~\cite{arun2021assessing,saporta2022benchmarking}.

Explanation-aware training incorporates explanation-related objectives directly into model optimization. Approaches include attribution regularization~\cite{ross2017right,rieger2020interpretations}, spatial supervision of attention or attribution maps such as GradMask and GAIN~\cite{simpson2019gradmask,li2018tell}, and weakly supervised localization based on image-level labels or pseudo-masks~\cite{kolesnikov2016seed,jung2023weakly,jo2021puzzle}. Depending on their formulation, these methods may require additional supervision, spatial annotations, architectural modifications, or domain-specific priors. CGC~\cite{pillai2022consistent} does not require pixel-level explanation supervision; instead, it applies contrastive learning to encourage agreement between Grad-CAM explanations of an image and its transformed counterpart. Spatial agreement with annotated regions is one criterion used to evaluate the resulting explanations, including in the original CGC study~\cite{pillai2022consistent}. Localization, perturbation-based faithfulness, and transformation consistency nevertheless quantify distinct properties and should not be treated as interchangeable.

\subsection{Localization and Model Reliance}
Localization metrics quantify spatial agreement between a saliency map and a reference region. Pointing Game (PG)~\cite{zhang2018top} records whether the maximally salient pixel lies within the reference region, whereas Relevance Mass Accuracy (RMA) measures the fraction of total saliency contained within that region~\cite{arras2022clevr}. Both metrics therefore measure agreement with a spatial annotation rather than dependence of the model output on the annotated region.

However, lesion overlap is not equivalent to faithfulness. A classifier may use both lesion and non-lesion evidence. Greater concentration of a heatmap within the lesion does not establish that the corresponding pixels are the most influential under a specified perturbation. This distinction is particularly relevant in medical imaging because acquisition and dataset-construction artifacts can themselves carry predictive information.

The Cheng brain tumor dataset used in our case study provides a concrete example of this distinction. Prior work showed that tumor type can be predicted from non-tumor information in this benchmark, including cues associated with representative-slice selection and anatomy~\cite{wallis2022cleverhans}. High classification accuracy on this dataset therefore does not establish that the classifier relies primarily on clinically relevant tumor information. Accordingly, improved lesion overlap supports a localization claim but does not, by itself, establish increased reliance on clinically relevant evidence.

\subsection{Protocol Dependence of Perturbation-Based Faithfulness Metrics}
Perturbation-based faithfulness metrics evaluate how the model output changes when input features are modified according to their attribution ranking. DAUC~\cite{petsiuk2018rise} progressively perturbs pixels from highest to lowest saliency while recording the target-class probability. IAUC~\cite{petsiuk2018rise} progressively restores pixels from a specified baseline image in the same saliency order while recording the target-class probability. IROF~\cite{rieger2020irof} partitions the input into regions, ranks them by mean attribution, and progressively removes them while recording the target-class probability.

The interpretation of these metrics is necessarily conditional on the perturbation protocol. Different baselines, replacement values, region definitions, or perturbation orders can produce different outcomes. Perturbations may also produce inputs outside the training distribution. We therefore regard these metrics as complementary evidence of model reliance rather than definitive measures of explanation quality~\cite{fel2022xplique,hedstrom2023quantus}.

\subsection{Direct Evaluation of the Targeted Explanation Property}

When an explanation-aware objective targets a particular property, that property should be evaluated directly. CGC, for example, explicitly encourages agreement between the heatmap of a transformed image and the correspondingly transformed heatmap of the original image~\cite{pillai2022consistent}. Localization and faithfulness are downstream properties; neither substitutes for a direct transformation-consistency measure.

Our case study includes qualitative examples under transformation but does not report a quantitative transformation-consistency metric. We therefore treat transformation consistency as quantitatively unmeasured and do not infer it from improved localization. More generally, an explanation-aware objective should not be evaluated solely through a different downstream property.

\section{Case Study: Contrastive Grad-CAM Consistency in Brain MRI}
\label{sec:case}

\subsection{CGC Formulation}

Let $\mathcal{D}=\{(x_i,\ell_i)\}_{i=1}^{N}$ denote a dataset of $N$
single-channel MRI slices $x_i$ and corresponding class labels $\ell_i$.
We use a ResNet-50~\cite{he2016deep} with one input channel and three output
classes. For an input image $x$ and target class $t$, Grad-CAM
~\cite{selvaraju2017grad} is computed from the activations and class-score
gradients of a selected convolutional layer. Let
$F^k\in\mathbb{R}^{H_f\times W_f}$ denote the $k$-th feature map of this
layer, with activation $F^k_{p,q}$ at spatial location $(p,q)$, and let
$y^t$ denote the pre-softmax logit for class $t$. The Grad-CAM weight
associated with feature map $k$ is
\begin{equation}
    \alpha_k^t =
    \frac{1}{H_fW_f}
    \sum_{p=1}^{H_f}\sum_{q=1}^{W_f}
    \frac{\partial y^t}{\partial F^k_{p,q}},
\end{equation}
and the corresponding class-specific heatmap is
\begin{equation}
    g^t(x)=
    \mathrm{ReLU}\!\left(
    \sum_k \alpha_k^t F^k
    \right).
\end{equation}
The ReLU retains the positive values of the weighted activation map.

Following the original CGC formulation~\cite{pillai2022consistent}, we use
$g_i(\cdot)$ to denote Grad-CAM computed with respect to the top predicted
class of the original, untransformed image $x_i$. This class is held fixed
when Grad-CAM is computed for the corresponding transformed image. For a
mini-batch $\{x_j\}_{j=1}^{n}$, each image is associated with an
independently sampled geometric transformation $T_j(\cdot)$. For query
image $x_i$, the transformed explanation of the original image,
$T_i(g_i(x_i))$, and the explanation of the transformed image,
$g_i(T_i(x_i))$, form the positive pair. The explanations
$\{g_j(T_j(x_j))\}_{j\neq i}$ form the negative pairs. The per-query CGC
loss is
\begin{equation}
    L_i =
    -\log
    \frac{
        \exp\!\left(
            \mathrm{sim}\!\left(
                T_i(g_i(x_i)),\,g_i(T_i(x_i))
            \right)/\tau
        \right)
    }{
        \sum_{j=1}^{n}
        \exp\!\left(
            \mathrm{sim}\!\left(
                T_i(g_i(x_i)),\,g_j(T_j(x_j))
            \right)/\tau
        \right)
    },
    \label{eq:cgc}
\end{equation}
where $\mathrm{sim}(\cdot,\cdot)$ denotes cosine similarity and $\tau>0$
is the temperature parameter.

With each image serving once as the query, the batch-level CGC loss is
\begin{equation}
    \mathcal{L}_{\mathrm{CGC}}
    =
    \sum_{i=1}^{n} L_i.
\end{equation}
The training objective is
\begin{equation}
    \mathcal{L}
    =
    \mathcal{L}_{\mathrm{CE}}
    +
    \lambda\mathcal{L}_{\mathrm{CGC}},
\end{equation}
where $\mathcal{L}_{\mathrm{CE}}$ is the cross-entropy classification loss
and $\lambda$ controls the contribution of the CGC term. Tumor masks are
not used in the training objective.

\subsection{Dataset and Experimental Design}

The Brain Tumor MRI dataset~\cite{Cheng2017brain} contains 3,064
contrast-enhanced T1-weighted MRI slices from 233 patients, with corresponding
binary tumor masks delineated by experienced radiologists. We use five-fold
stratified patient-grouped cross-validation, with all slices from each patient
assigned to a single fold. Within each training partition, a second
patient-grouped split reserves approximately $10\%$ of the training data for
validation. Across folds, the training splits contain 2,147--2,270 images from
166--170 patients, the validation splits contain 225--278 images from 18--20
patients, and the test splits contain 531--687 images from 45--48 patients.

The baseline and CGC models use the same ResNet-50 architecture and identical
patient-level data splits. Models are trained for 20 epochs using stochastic
gradient descent with an initial learning rate of $0.01$, momentum $0.9$,
weight decay $10^{-4}$, and batch size 16. The learning rate is reduced by a
factor of $0.1$ every five epochs. The model input resolution is
$384\times384$ pixels. Training augmentation consists of random resized
cropping with scale range $(0.08,1.0)$, followed by random horizontal
flipping. CGC uses $\lambda=0.25$ and $\tau=0.5$. Grad-CAM is computed from
the final ResNet-50 residual stage, \texttt{layer4}. All experiments are run
with a fixed random seed of 42.

The crop scale range is broad for brain MRI and may exclude clinically
relevant anatomy. We therefore do not assume that all sampled transformations
preserve all clinically relevant anatomy, and we interpret the qualitative
transformation examples accordingly.

\subsection{Evaluation}

We report classification accuracy as the predictive-performance measure.
Explanations are evaluated along two distinct dimensions: lesion localization
and perturbation-based faithfulness. Localization is quantified using Pointing
Game (PG)~\cite{zhang2018top} and Relevance Mass Accuracy
(RMA)~\cite{arras2022clevr}. Perturbation-based faithfulness is evaluated
using Insertion AUC (IAUC) and Deletion AUC
(DAUC)~\cite{petsiuk2018rise}, together with Iterative Removal Of Features
(IROF)~\cite{rieger2020irof}. Higher values are preferred for PG, RMA, IAUC,
and IROF, whereas lower values are preferred for DAUC. IAUC, DAUC, and IROF
are computed using the ExplainableAIImageMeasures
implementation~\cite{meier2021ExplainableAIImageMeasures}, with 70
superpixels used for IROF. Results are reported as mean $\pm$ standard
deviation across the five patient-grouped folds. We interpret these fold-level
statistics descriptively and make no inferential claims from small differences
in mean performance.

\section{Empirical Results}
\label{sec:evidence}

\subsection{Classification and Localization}

Table~\ref{tab:main_results} reports classification accuracy and Grad-CAM
localization results. Overall classification accuracy changes from
$93.63\pm1.91\%$ for the baseline to $94.87\pm1.56\%$ with CGC. We report
this 1.24-percentage-point difference descriptively and do not interpret it
as evidence of improved classification performance.

Overall PG changes from $27.24\pm3.76\%$ to $36.02\pm4.00\%$, an absolute
difference of 8.78 percentage points. RMA changes from $11.17\pm1.60\%$ to
$18.01\pm2.70\%$, an absolute difference of 6.84 percentage points. Mean PG
and RMA are higher with CGC for all three tumor classes.

\begin{table}[t]
\centering
\small
\caption{Classification accuracy and Grad-CAM localization. Values are mean
$\pm$ standard deviation across five patient-grouped folds. Higher values are
preferred.}
\label{tab:main_results}
\begin{tabular}{|l|c|c|c|c|c|c|}
\hline
& \multicolumn{2}{c|}{Accuracy (\%)} &
\multicolumn{2}{c|}{PG (\%)} &
\multicolumn{2}{c|}{RMA (\%)} \\
\hline
Class & Baseline & CGC & Baseline & CGC & Baseline & CGC \\
\hline
Meningioma & $88.10\!\pm\!7.70$ & $90.82\!\pm\!7.50$ &
$57.27\!\pm\!8.35$ & $75.61\!\pm\!6.43$ &
$22.25\!\pm\!4.53$ & $36.03\!\pm\!4.43$ \\
Glioma & $95.78\!\pm\!1.30$ & $97.47\!\pm\!1.90$ &
$24.98\!\pm\!7.02$ & $31.29\!\pm\!8.72$ &
$10.45\!\pm\!1.47$ & $17.08\!\pm\!3.08$ \\
Pituitary & $94.87\!\pm\!5.90$ & $94.26\!\pm\!5.20$ &
$7.13\!\pm\!3.21$ & $12.67\!\pm\!7.16$ &
$3.21\!\pm\!0.61$ & $5.11\!\pm\!1.35$ \\
\hline
Overall & $93.63\!\pm\!1.91$ & $\mathbf{94.87\!\pm\!1.56}$ &
$27.24\!\pm\!3.76$ & $\mathbf{36.02\!\pm\!4.00}$ &
$11.17\!\pm\!1.60$ & $\mathbf{18.01\!\pm\!2.70}$ \\
\hline
\end{tabular}
\end{table}

For pituitary tumors, classification accuracy remains above $94\%$ for both
models, whereas PG remains below $13\%$ and RMA below $6\%$. High
classification accuracy therefore coexists with weak Grad-CAM localization
relative to the annotated tumor. These results do not establish whether
non-tumor features contribute to the predictions, but they demonstrate that
classification performance and lesion localization provide distinct
information.

\subsection{Perturbation-Based Faithfulness}

Table~\ref{tab:faithfulness} reports the perturbation-based faithfulness
results. Overall IAUC is higher with CGC ($0.78\pm0.07$ vs.\
$0.73\pm0.06$), whereas DAUC favors the baseline ($0.36\pm0.07$ vs.\
$0.41\pm0.07$; lower is better). Overall IROF is $0.57$ for both models at
the reported precision. The class-wise comparisons likewise show no
consistent direction across metrics.

\begin{table}[t]
\centering
\small
\caption{Perturbation-based faithfulness. Values are mean $\pm$ standard
deviation across five patient-grouped folds. Higher values are preferred for
IAUC and IROF; lower values are preferred for DAUC.}
\label{tab:faithfulness}
\begin{tabular}{|l|l|c|c|c|c|}
\hline
Metric & Model & Meningioma & Glioma & Pituitary & Overall \\
\hline
IAUC ($\uparrow$) & Baseline & $0.96\pm0.03$ & $0.81\pm0.06$ &
$0.43\pm0.12$ & $0.73\pm0.06$ \\
& CGC & $0.90\pm0.12$ & $0.92\pm0.09$ &
$0.44\pm0.19$ & $\mathbf{0.78\pm0.07}$ \\
\hline
DAUC ($\downarrow$) & Baseline & $0.69\pm0.13$ & $0.39\pm0.13$ &
$0.07\pm0.05$ & $\mathbf{0.36\pm0.07}$ \\
& CGC & $0.41\pm0.23$ & $0.62\pm0.21$ &
$0.06\pm0.04$ & $0.41\pm0.07$ \\
\hline
IROF ($\uparrow$) & Baseline & $0.44\pm0.05$ & $0.42\pm0.06$ &
$0.91\pm0.03$ & $\mathbf{0.57\pm0.04}$ \\
& CGC & $0.53\pm0.15$ & $0.39\pm0.16$ &
$0.92\pm0.03$ & $\mathbf{0.57\pm0.07}$ \\
\hline
\end{tabular}
\end{table}

Taken together, the results support a narrower conclusion: CGC yields higher
mean lesion-localization scores in this benchmark, whereas the selected
perturbation-based metrics do not indicate a consistent faithfulness gain.
This separation supports our position that higher localization scores should
not be interpreted as evidence of a general improvement in faithfulness.

\subsection{Qualitative Transformation Examples}

\begin{figure}[t]
    \centering
    \includegraphics[width=\textwidth]{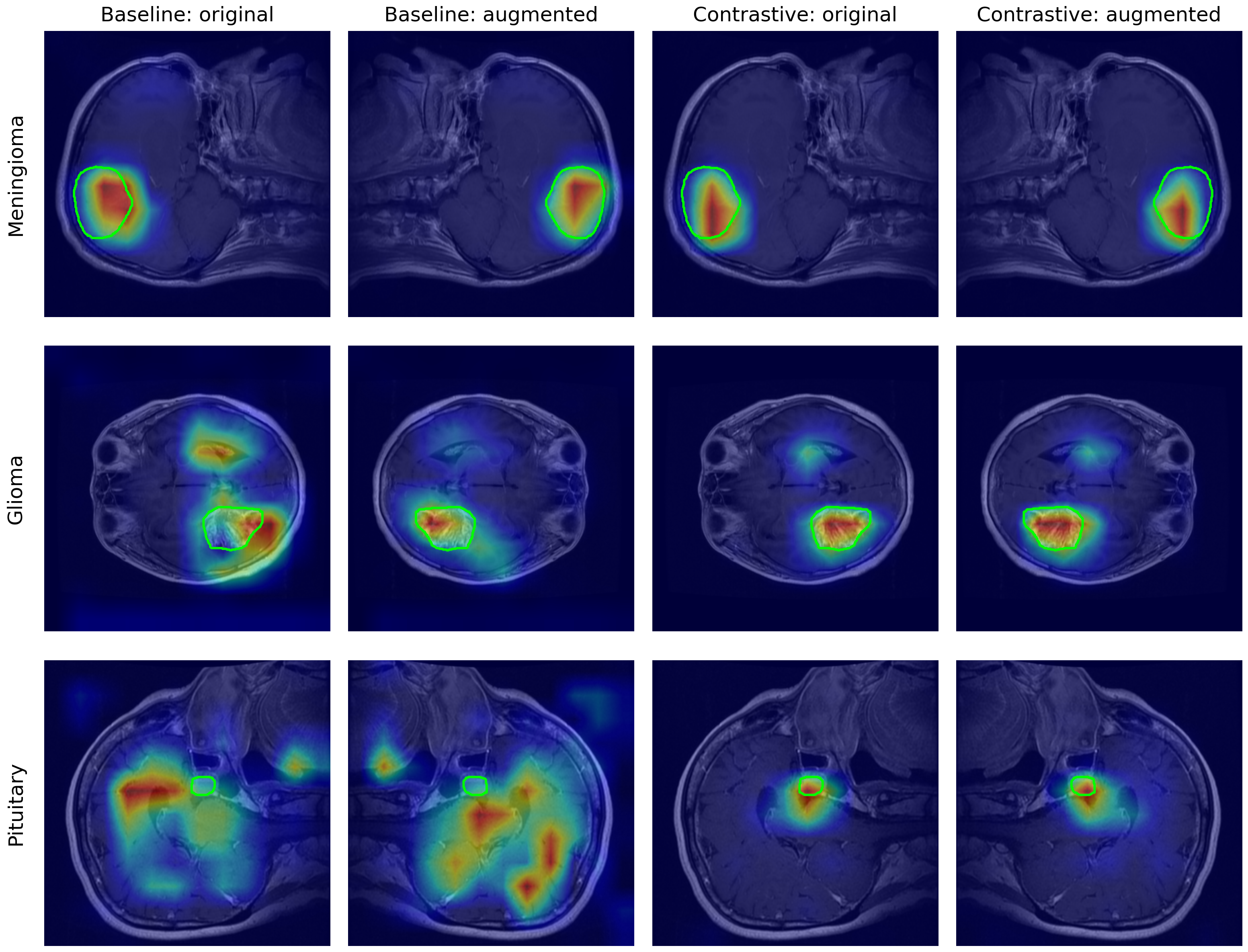}
    \caption{Selected Grad-CAM explanations for baseline and CGC models on
    meningioma, glioma, and pituitary test images. Columns show original and
    horizontally flipped inputs for each model; green contours denote
    ground-truth tumor masks. The examples are shown for qualitative
    comparison and are not used to quantify transformation consistency.}
    \label{fig:qualitative_consistency}
\end{figure}

Figure~\ref{fig:qualitative_consistency} shows selected Grad-CAM maps for
original and horizontally flipped images. In these examples, the CGC maps
appear more spatially similar between the original and horizontally flipped
inputs and more concentrated in the annotated tumor region. These examples
are qualitatively consistent with the behavior targeted by CGC but do not
provide a quantitative measure of transformation consistency.

Quantitative assessment of transformation consistency requires a direct
consistency metric. Our results therefore support the reported localization
and perturbation-based comparisons, but do not quantify transformation
consistency.

\section{Implications for XAI Evaluation}
\label{sec:implications}

Our empirical results lead to three methodological implications for the
evaluation of explanation-aware training.

\paragraph{Property-specific interpretation.}
Improvement within a single metric family supports a claim about the
corresponding measured property rather than explanation quality as a whole.
Higher PG and RMA indicate stronger spatial agreement with the annotated region
according to these localization criteria. Changes in IAUC or DAUC characterize
model behavior under the corresponding insertion or deletion protocol. Claims
should therefore remain aligned with the property and protocol being measured.

\paragraph{Direct evaluation of targeted properties.}
An explanation-aware objective that targets a specific property should be
evaluated directly on that property. For CGC, transformation consistency should
therefore be quantified directly in addition to localization and
perturbation-based faithfulness. More generally, evaluation should include the
property explicitly targeted by the training objective.

\paragraph{Interpretation of metric disagreement.}
Disagreement across metrics should be reported rather than collapsed into a
single favorable conclusion. In our case study, IAUC, DAUC, and IROF rank the
two models differently, so the higher localization scores cannot be interpreted
as a uniform improvement in model reliance. Metric disagreement is therefore
informative about the scope of the observed differences.

\section{Limitations and Scope}
\label{sec:limitations}

The case study uses one 2-D MRI dataset, one CNN backbone, one attribution
method, and one fixed random seed. The five patient-grouped folds reflect
variation across patient-level splits but do not characterize variability
across training seeds. The broad crop scale range may exclude clinically
relevant anatomy in some transformed images. Transformation behavior is
examined qualitatively, without a quantitative measure of transformation
consistency. IAUC, DAUC, and IROF are conditional on their respective
perturbation protocols.

A documented bias in the Cheng dataset permits tumor classification from
non-tumor information~\cite{wallis2022cleverhans}. Accordingly, we do not
interpret the reported classification accuracy as evidence of reliance on
clinically relevant tumor information. Spatial agreement with tumor
annotations therefore remains distinct from evidence of model reliance.

Our results do not establish that CGC is generally preferable for medical
imaging. Our results instead support the more limited conclusion that
localization and perturbation-based faithfulness should be evaluated and
reported separately.

\section{Conclusion}
\label{sec:conclusion}

We argue that transformation consistency, localization, and faithfulness
should be treated as distinct evaluation targets. Our brain MRI case study
using CGC provides empirical support for this distinction. Mean PG changes
from $27.24\%$ for the baseline to $36.02\%$ with CGC, and mean RMA from
$11.17\%$ to $18.01\%$. The perturbation-based metrics show different
overall patterns: IAUC is higher with CGC, DAUC is lower for the baseline,
and IROF is unchanged at the reported precision.

CGC yields higher mean lesion-localization scores in this benchmark, whereas
the selected perturbation-based faithfulness metrics do not show a consistent
improvement. We therefore argue that XAI studies should directly evaluate the
property targeted by an explanation-aware objective, report spatial agreement
and model reliance as separate evaluation criteria, and explicitly report
disagreement across metrics. This distinction is particularly important in
medical imaging, where visually plausible explanations do not by themselves
establish reliance on clinically relevant evidence.

Future work will explore whether the difference between localization and faithfulness remains consistent across other medical imaging datasets, model architectures, and attribution methods. Alternative formulations of the CGC objective, including distribution-alignment approaches~\cite{chen2024your} and optimal transport, could be explored as well.

\begin{credits}
\subsubsection{\ackname}
The authors acknowledge the support of the French Agence Nationale de la
Recherche (ANR), under grant ANR-25-CE45-1573 (project EQUR-XSM).
All computations presented in this paper were performed using the GRICAD
infrastructure (\url{https://gricad.univ-grenoble-alpes.fr}), which is
supported by the Grenoble research community.

\end{credits}

\bibliographystyle{splncs04}
\bibliography{references}

@article{arun2021assessing,
title={Assessing the trustworthiness of saliency maps for localizing abnormalities in medical imaging},
author={Arun, Nishanth and Gaw, Nathan and Singh, Praveer and Chang, Ken and Aggarwal, Mehak and Chen, Bryan and Hoebel, Katharina and Gupta, Sharut and Patel, Jay and Gidwani, Mishka and others},
journal={Radiology: Artificial Intelligence},
volume={3},
number={6},
pages={e200267},
year={2021},
publisher={Radiological Society of North America}
}

@article{adebayo2018sanity,
title={Sanity checks for saliency maps},
author={Adebayo, Julius and Gilmer, Justin and Muelly, Michael and Goodfellow, Ian and Hardt, Moritz and Kim, Been},
journal={Advances in neural information processing systems},
volume={31},
year={2018}
}

@article{saporta2022benchmarking,
title={Benchmarking saliency methods for chest X-ray interpretation},
author={Saporta, Adriel and Gui, Xiaotong and Agrawal, Ashwin and Pareek, Anuj and Truong, Steven QH and Nguyen, Chanh DT and Ngo, Van-Doan and Seekins, Jayne and Blankenberg, Francis G and Ng, Andrew Y},
journal={Nature Machine Intelligence},
volume={4},
number={10},
pages={867--878},
year={2022},
publisher={Nature Publishing Group UK London}
}

@inproceedings{zhou2016CAM,
title={Learning deep features for discriminative localization},
author={Zhou, Bolei and Khosla, Aditya and Lapedriza, Agata and Oliva, Aude and Torralba, Antonio},
booktitle={Proceedings of the IEEE conference on computer vision and pattern recognition},
pages={2921--2929},
year={2016}
}

@inproceedings{li2018tell,
title={Tell me where to look: Guided attention inference network},
author={Li, Kunpeng and Wu, Ziyan and Peng, Kuan-Chuan and Ernst, Jan and Fu, Yun},
booktitle={Proceedings of the IEEE conference on computer vision and pattern recognition},
pages={9215--9223},
year={2018}
}

@inproceedings{kolesnikov2016seed,
title={Seed, expand and constrain: Three principles for weakly-supervised image segmentation},
author={Kolesnikov, Alexander and Lampert, Christoph H},
booktitle={European conference on computer vision},
pages={695--711},
year={2016},
organization={Springer}
}

@inproceedings{jo2021puzzle,
title={Puzzle-cam: Improved localization via matching partial and full features},
author={Jo, Sanghyun and Yu, In-Jae},
booktitle={2021 IEEE international conference on image processing (ICIP)},
pages={639--643},
year={2021},
organization={IEEE}
}

@article{jung2023weakly,
title={Weakly supervised thoracic disease localization via disease masks},
author={Jung, Hong-Gyu and Nam, Woo-Jeoung and Kim, Hyun-Woo and Lee, Seong-Whan},
journal={Neurocomputing},
volume={517},
pages={34--43},
year={2023},
publisher={Elsevier}
}

@misc{Cheng2017brain,
author       = {Junyao Cheng},
title        = {Brain magnetic resonance imaging tumor dataset},
year         = {2017},
howpublished = {Figshare, MRI Dataset, Version 8},
note         = {Dataset},
}

@inproceedings{he2016deep,
title={Deep Residual Learning for Image Recognition},
author={He, Kaiming and Zhang, Xiangyu and Ren, Shaoqing and Sun, Jian},
booktitle={Proceedings of the IEEE Conference on Computer Vision and Pattern Recognition},
pages={770--778},
year={2016}
}

@inproceedings{ross2017right,
title={Right for the Right Reasons: Training Differentiable Models by Constraining their Explanations},
author={Ross, Andrew S and Hughes, Michael C and Doshi-Velez, Finale},
booktitle={Proceedings of the Twenty-Sixth IJCAI},
year={2017}
}

@inproceedings{rieger2020interpretations,
title={Interpretations are useful: penalizing explanations to align neural networks with prior knowledge},
author={Rieger, Laura and Singh, Chandan and Murdoch, William and Yu, Bin},
booktitle={International conference on machine learning},
pages={8116--8126},
year={2020},
organization={PMLR}
}

@article{simpson2019gradmask,
title={GradMask: Reduce Overfitting by Regularizing Saliency},
author={Simpson, Becks and Dutil, Francis and Bengio, Yoshua and Cohen, Joseph Paul},
journal={ArXiv},
year={2019}
}

@inproceedings{fong2017meaningful,
title={Interpretable Explanations of Black Boxes by Meaningful Perturbation},
author={Fong, Ruth and Vedaldi, Andrea},
booktitle={Proceedings of the ICCV},
year={2017}
}

@inproceedings{selvaraju2017grad,
title={Grad-cam: Visual explanations from deep networks via gradient-based localization},
author={Selvaraju, Ramprasaath R and Cogswell, Michael and Das, Abhishek and Vedantam, Ramakrishna and Parikh, Devi and Batra, Dhruv},
booktitle={Proceedings of the IEEE international conference on computer vision},
pages={618--626},
year={2017}
}

@inproceedings{chattopadhay2018gradcamp,
title={Grad-cam++: Generalized gradient-based visual explanations for deep convolutional networks},
author={Chattopadhay, Aditya and Sarkar, Anirban and Howlader, Prantik and Balasubramanian, Vineeth N},
booktitle={2018 IEEE winter conference on applications of computer vision (WACV)},
pages={839--847},
year={2018},
organization={IEEE}
}

@article{jiang2021layercam,
title={LayerCAM: Exploring Hierarchical Class Activation Maps for Localization},
author={Jiang, Peng-Tao and Zhang, Chang-Bin and Hou, Qibin and Cheng, Ming-Ming and Wei, Yunchao},
journal={IEEE Transactions on Image Processing},
volume={30},
pages={5875--5888},
year={2021},
doi={10.1109/TIP.2021.3089943}
}

@inproceedings{wang2020score,
title={Score-CAM: Score-weighted visual explanations for convolutional neural networks},
author={Wang, Haofan and Wang, Zifan and Du, Mengnan and Yang, Fan and Zhang, Zijian and Ding, Sirui and Mardziel, Piotr and Hu, Xia},
booktitle={Proceedings of the IEEE/CVF conference on computer vision and pattern recognition workshops},
pages={24--25},
year={2020}
}

@article{banymuhammad2021eigencam,
title={Eigen-CAM: Visual Explanations for Deep Convolutional Neural Networks},
author={Bany Muhammad, Mohammed and Yeasin, Mohammed},
journal={SN Computer Science},
volume={2},
number={1},
year={2021},
doi={10.1007/s42979-021-00449-3}
}

@inproceedings{pillai2022consistent,
title={Consistent explanations by contrastive learning},
author={Pillai, Vipin and Koohpayegani, Soroush Abbasi and Ouligian, Ashley and Fong, Dennis and Pirsiavash, Hamed},
booktitle={Proceedings of the IEEE/CVF Conference on Computer Vision and Pattern Recognition},
pages={10213--10222},
year={2022}
}

@article{chen2024your,
title={Your contrastive learning problem is secretly a distribution alignment problem},
author={Chen, Zihao and Lin, Chi-Heng and Liu, Ran and Xiao, Jingyun and Dyer, Eva L},
journal={Advances in Neural Information Processing Systems},
volume={37},
pages={91597--91617},
year={2024}
}

@article{petsiuk2018rise,
title={Rise: Randomized input sampling for explanation of black-box models},
author={Petsiuk, Vitali and Das, Abir and Saenko, Kate},
journal={arXiv preprint arXiv:1806.07421},
year={2018}
}

@article{rieger2020irof,
title={Irof: a low resource evaluation metric for explanation methods},
author={Rieger, Laura and Hansen, Lars Kai},
journal={arXiv preprint arXiv:2003.08747},
year={2020}
}

@article{zhang2018top,
title={Top-down neural attention by excitation backprop},
author={Zhang, Jianming and Bargal, Sarah Adel and Lin, Zhe and Brandt, Jonathan and Shen, Xiaohui and Sclaroff, Stan},
journal={International Journal of Computer Vision},
volume={126},
number={10},
pages={1084--1102},
year={2018},
publisher={Springer}
}

@article{arras2022clevr,
title={CLEVR-XAI: A benchmark dataset for the ground truth evaluation of neural network explanations},
author={Arras, Leila and Osman, Ahmed and Samek, Wojciech},
journal={Information Fusion},
volume={81},
pages={14--40},
year={2022},
publisher={Elsevier}
}

@article{hedstrom2023quantus,
title={Quantus: An explainable ai toolkit for responsible evaluation of neural network explanations and beyond},
author={Hedstr{"o}m, Anna and Weber, Leander and Krakowczyk, Daniel and Bareeva, Dilyara and Motzkus, Franz and Samek, Wojciech and Lapuschkin, Sebastian and H{"o}hne, Marina M-C},
journal={Journal of Machine Learning Research},
volume={24},
number={34},
pages={1--11},
year={2023}
}

@misc{meier2021ExplainableAIImageMeasures,
title        = {ExplainableAIImageMeasures},
author       = {Meier, Johannes},
year         = {2021},
url          = {https://github.com/meier-johannes94/ExplainableAIImageMeasures},
}

@article{fel2022xplique,
title={Xplique: A deep learning explainability toolbox},
author={Fel, Thomas and Hervier, Lucas and Vigouroux, David and Poche, Antonin and Plakoo, Justin and Cadene, Remi and Chalvidal, Mathieu and Colin, Julien and Boissin, Thibaut and Bethune, Louis and others},
journal={arXiv preprint arXiv:2206.04394},
year={2022}
}

@article{simonyan2013deep,
title={Deep Inside Convolutional Networks: Visualising Image Classification Models and Saliency Maps},
author={Simonyan, Karen and Vedaldi, Andrea and Zisserman, Andrew},
journal={arXiv preprint arXiv:1312.6034},
year={2013}
}

@inproceedings{sundararajan2017axiomatic,
title={Axiomatic Attribution for Deep Networks},
author={Sundararajan, Mukund and Taly, Ankur and Yan, Qiqi},
booktitle={Proceedings of the 34th International Conference on Machine Learning (ICML)},
volume={70},
pages={3319--3328},
year={2017},
organization={PMLR}
}

@article{wallis2022cleverhans,
  author  = {Wallis, David and Buvat, Ir{\`e}ne},
  title   = {Clever Hans effect found in a widely used brain tumour MRI dataset},
  journal = {Medical Image Analysis},
  volume  = {77},
  pages   = {102368},
  year    = {2022},
  doi     = {10.1016/j.media.2022.102368}
}

\end{document}